\documentclass{article}

\usepackage{microtype}
\usepackage{graphicx}
\usepackage{subfigure}
\usepackage{booktabs} 

\usepackage{hyperref}

\usepackage[accepted]{icml2020}

\icmltitlerunning{Lanfrica}

\begin{document}

\twocolumn[
\icmltitle{Lanfrica: A Participatory Approach to Documenting Machine Translation Research on African Languages}



\icmlsetsymbol{equal}{*}

\begin{icmlauthorlist}
\icmlauthor{Chris C. Emezue}{to}
\icmlauthor{Bonaventure F.P. Dossou}{goo}

\end{icmlauthorlist}

\icmlaffiliation{to}{Mathematics and Data Science, Technical University of Munich}
\icmlaffiliation{goo}{Data Engineering, Jacobs University}

\icmlcorrespondingauthor{}{}

\icmlkeywords{machine learning,lanfrica, african languages, database,low-resource}

\vskip 0.3in
]



\printAffiliationsAndNotice{\icmlEqualContribution} 
\begin{abstract}
Over the years, there have been campaigns to include the African languages in the growing research on machine translation (MT) in particular, and natural language processing (NLP) in general. Africa has the highest language diversity, with 1500-2000 documented languages and many more undocumented or extinct languages\cite{ethnologue,africa}. This makes it hard to keep track of the MT research, models and dataset that have been developed for some of them. As the internet and social media make up the daily lives of more than half of the world\cite{lin_2020}, as well as over 40\% of Africans\cite{campbell_2019}, online platforms can be useful in creating accessibility to researches, benchmarks and datasets in these African languages, thereby improving reproducibility and sharing of existing research and their results. In this paper, we introduce Lanfrica, a novel, on-going framework that employs a participatory approach to documenting researches, projects, benchmarks and dataset on African languages.
\end{abstract}
\section{Background and Motivation}
African languages have received (and still receive) a great deal of attention in natural language processing and machine translation research in the last decade, with the advent of online communities and organizations like:
\begin{itemize}
\item Masakhane\footnote{\url{https://www.masakhane.io/home}}, an online, community-led, open-source research effort aimed at building and facilitating a community of NLP researchers for African languages.
\item Deep Learning Indaba\footnote{\url{https://deeplearningindaba.com/2020/}}, an organisation whose mission is to strengthen machine learning and artificial intelligence in Africa through. They work towards the goal of Africans being not only observers and receivers of the ongoing advances in AI, but also active shapers and owners of these technological advances.
\item BlackinAI\footnote{\url{https://blackinai.github.io/}}, a multi-institutional, transcontinental initiative creating a space for sharing ideas, fostering collaborations, and discussing initiatives to increase the presence of Black individuals in the field of AI. 
\item Zindi\footnote{\url{https://zindi.africa/}}, the first data science competition platform in Africa. With around 12000 registered users\cite{bright_2020}, Zindi gives organisations and governments access to world-class machine learning and AI solutions, as well as gives African data scientists a place to learn new skills, grow and access work opportunities. 
\end{itemize}
All these organizations, and many more unmentioned communities, are dedicated towards promoting AI research in Africa including machine translation and natural language processing in African languages (hereinafter called AfricaNLP). 

Major companies are also taking actions to improve AI diversity, reduce bias and include African languages in their technologies. In April, Google opened its first African AI research centre in Ghana\cite{russon_2019} to develop solutions to help improve healthcare, agriculture and education. Also, in a move to create more opportunities for the voices of more African researchers to be heard, the 2020 International Conference on Learning Representations (ICLR) conference was scheduled to take place in Addis Ababa, Ethiopia \cite{news}. 

All these attest to the fact that there is a movement now, more than ever, to shhare and showcase machine learning researches on African languages and put Africa on the online map in order to, among other things, preserve the African culture in our fast-growing digital world.

\cite{jade,jade2}, among other MT researchers on AfricaNLP, have shown that the major problems facing AfricaNLP are:
\begin{itemize}
\item lack of hope from the African society on African languages being used as a major mode of communication in the future
\item lack of resources for African languages 
\item \textit{\textbf{low discoverability of existing researchers on African languages}} \item \textit{\textbf{lack of publicly-available benchmarks}}
\item \textit{\textbf{low sharing of existing research and code}}
\end{itemize}
 
\section{Lanfrica}
Lanfrica is an online, community-led effort to specifically tackle the problems of low-discoverability, lack of publicly-available benchmarks and low sharing of existing researches on all African languages by:
\begin{itemize}
    \item creating an online, open-source database system for easy, at-a-glance access to existing NLP researches, MT research and results (including benchmarks) involving African languages.
    \item implementing a participatory, community-led approach to populating the database with existing researches on machine translation involving African languages. The participatory approach is explained more in section \ref{walk}.
\end{itemize} 
The main objective of Lanfrica is to document existing researches, research-results, benchmarks and projects (completed and ongoing) on African languages and present them in a user-friendly manner for immediate access by anyone. The major focus of the pilot stage of Lanfrica is machine translation.
\subsection{Participatory Research Approach}
\begin{figure}[!]
  \centering
    \includegraphics[width=\linewidth]{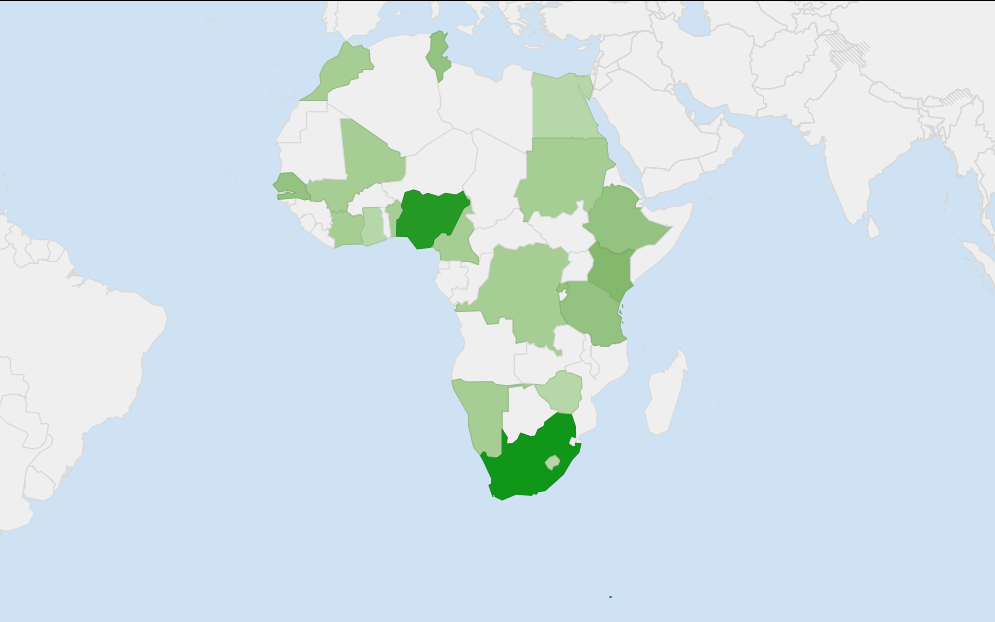}
 \caption{A map of Africa showing the participant countries of Masakhane. Source: Masakhane}
\label{africalang} 
\end{figure}

The work on Lanfrica relied on participant observations, analysis and experiments carried out while working with the Masakhane community. The Masakhane community consists of 144 participants from 17 African countries with diverse educations and occupations, and 2 countries outside Africa (USA and Germany)\ref{africalang}. As of February 2020, over 35 translation results for over 29 African languages have been published by over 25 contributors on GitHub.

Through analysis from surveys and interviews conducted by the authors, as well as collaborative work conducted with the help of the Masakhane community, it was found that a major bottleneck of reproducibility and smooth comparision of MT research (including benchmarks and information on testing data) on the several African languages is the lack of a publicly available system that documents such research and can easily be used to check for existing research and benchmarks.

For example, \textit{Researcher A} wants to work on building a neural machine translation (NMT) model for Fon language (particularly Fon-French). Thanks to the Masakhane community, \textit{Researcher A} is able to meet some (not all) researchers working on the Fon language, who may have already designed a Fon-French MT model. However, \textit{Researcher A} has to go through strenuous efforts, like organizing interviews/chats with such researchers in order to know their benchmark results, searching for their research paper (if published) or finding out the testing data for model comparison, all leading to much needed data on \textit{previous/related work on Fon-French translation}. The easiest option for \textit{Researcher A}, which is the common trend in this case, is to go on building the Fon-French NMT model, with little or no consideration of existing NMT models and their results, and try to publish a research paper on the obtained results, hoping that future researchers will be able to access the paper and use the results attained. 

The Masakhane community provides a \textit{Jupyter Notebook} which runs on Google Colab, contains models  built using Joey NMT\cite{DBLP:journals/corr/abs-1907-12484} and features documented data preparation, model configuration, training and evaluation\cite{Orife2020MasakhaneM} with the aim of improving reproducibility and discoverability. However, researchers working independently, or who have not been able to join Masakhane or use its services still face the same problem as \textit{Researcher A}.
\begin{figure}[!]
  \centering
    \includegraphics[width=\linewidth]{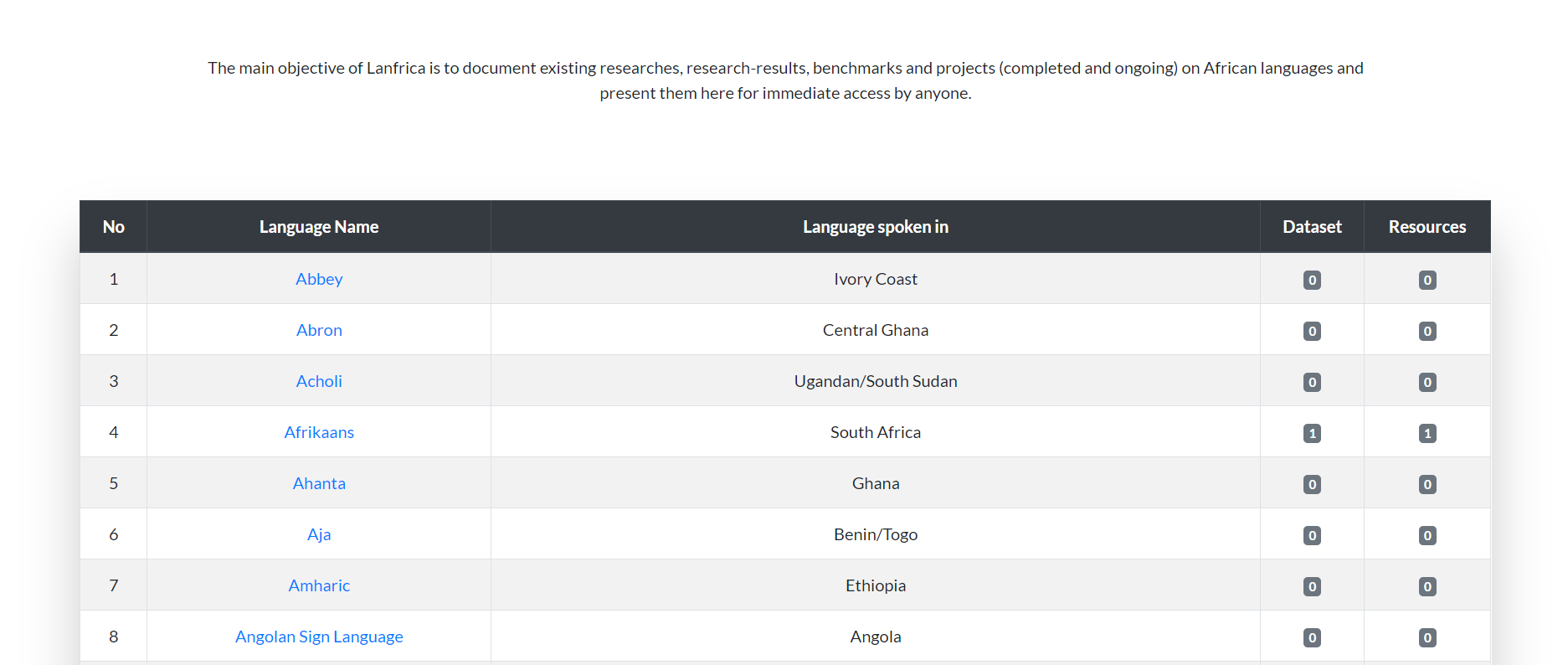}
 \caption{Part of landing page showing African languages and their number of available online dataset and resources }
\label{landing} 
\end{figure}
\begin{figure}
  \centering
    \includegraphics[width=\linewidth]{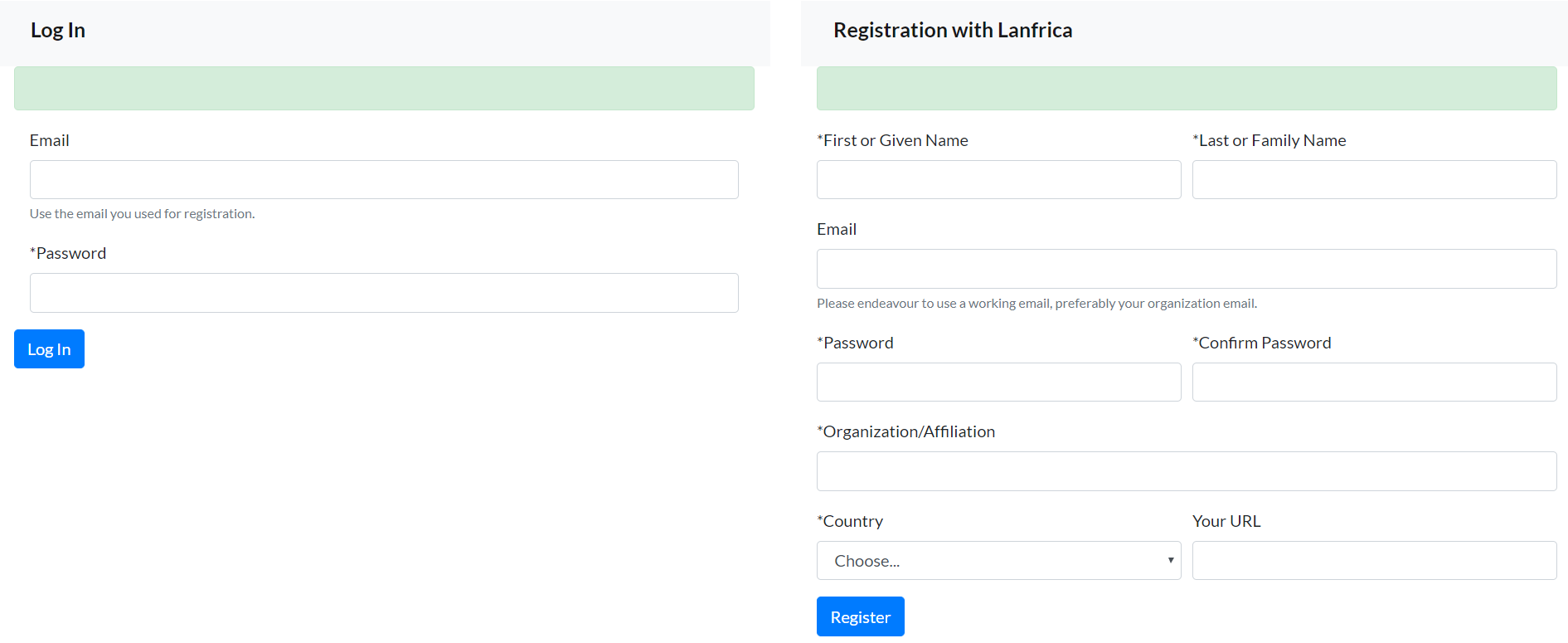}
 \caption{\small Log in/Create account page (still under construction)}
\label{create} 
\end{figure}

\subsection{Participatory Design Method}
\label{walk}

Lanfrica is going to be hosted online, as pre-analysis carried out by the authors shows that the platform will have greatest participation when hosted on a website. The walkthrough below serves to give a mental picture of how Lanfrica operates.

\begin{figure*}[!ht]
  \centering
    \includegraphics[width=\linewidth]{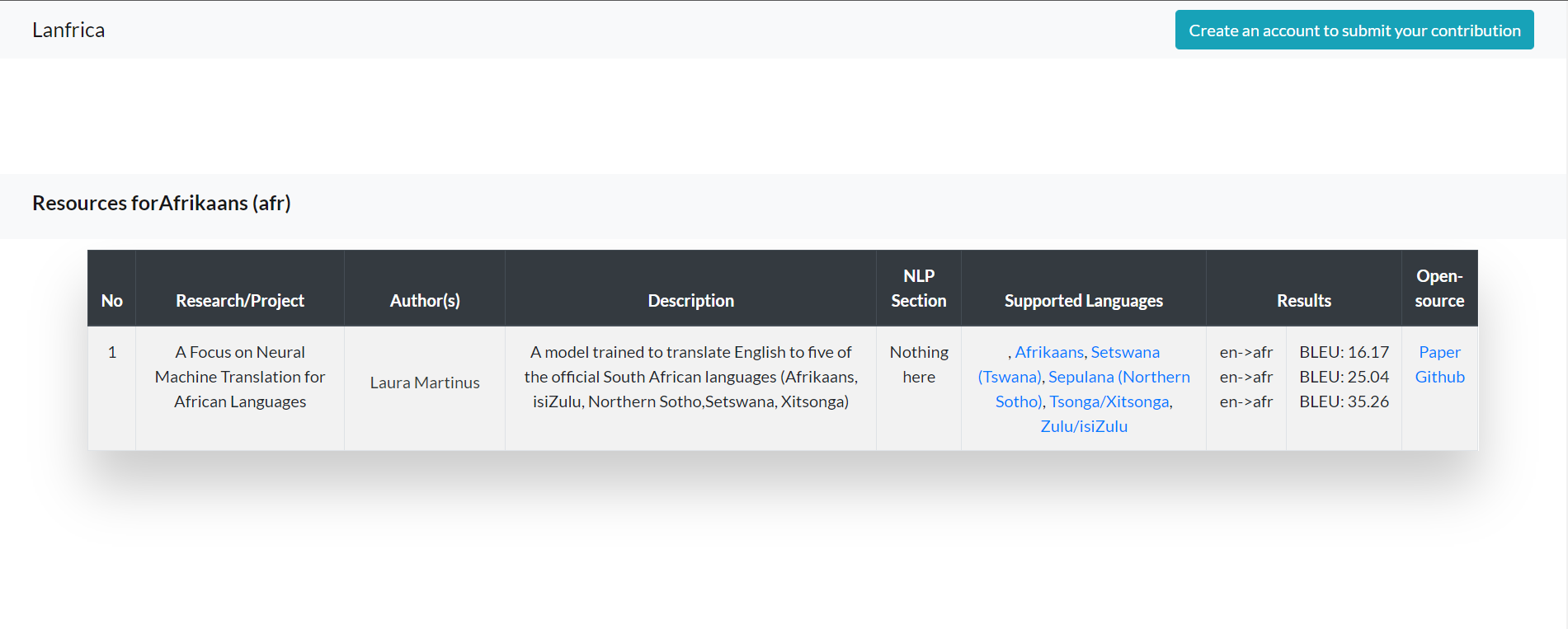}
 \caption{Page showing details of research/projects in Afrikaans (used as an example) }
\label{resources} 
\vspace{0.00mm} 
    \includegraphics[width=\linewidth]{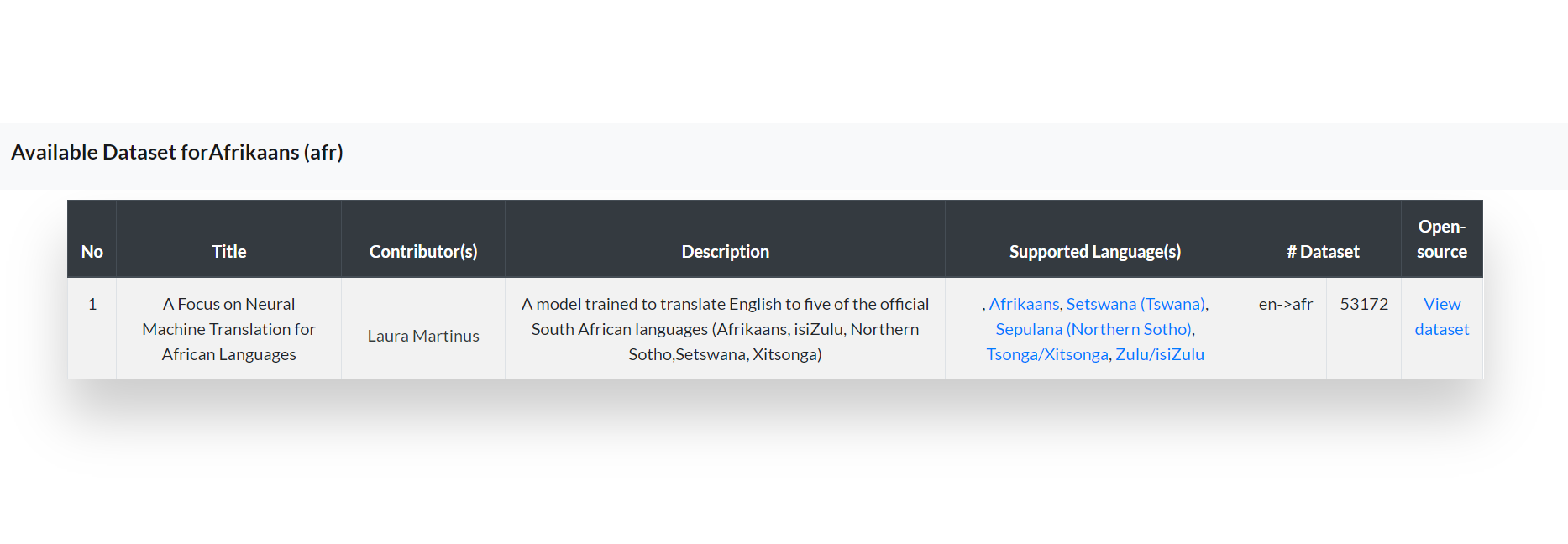}
 \caption{\small Webpage showing available dataset for Afrikaans (used as an example) }
\label{data} 
\end{figure*}

\textbf{Walkthrough:}
\begin{enumerate}
\item A user visits the Lanfrica website.
\item The user sees the landing page (see Figure ~\ref{landing}), showing a list of African languages (referenced with hyperlinks) and the options to:
\begin{enumerate}
\label{steps}
\item \label{stepa} Submit a contribution: in this case, the user is an author/co-author of a research paper or working (or co-working) on an MT project (completed or ongoing) and wants to store information about the research or project on Lanfrica. The user will then be requested to fill a form, which will contain information about the research/project.
\begin{figure}[!h]
  \centering
    \includegraphics[width=\linewidth]{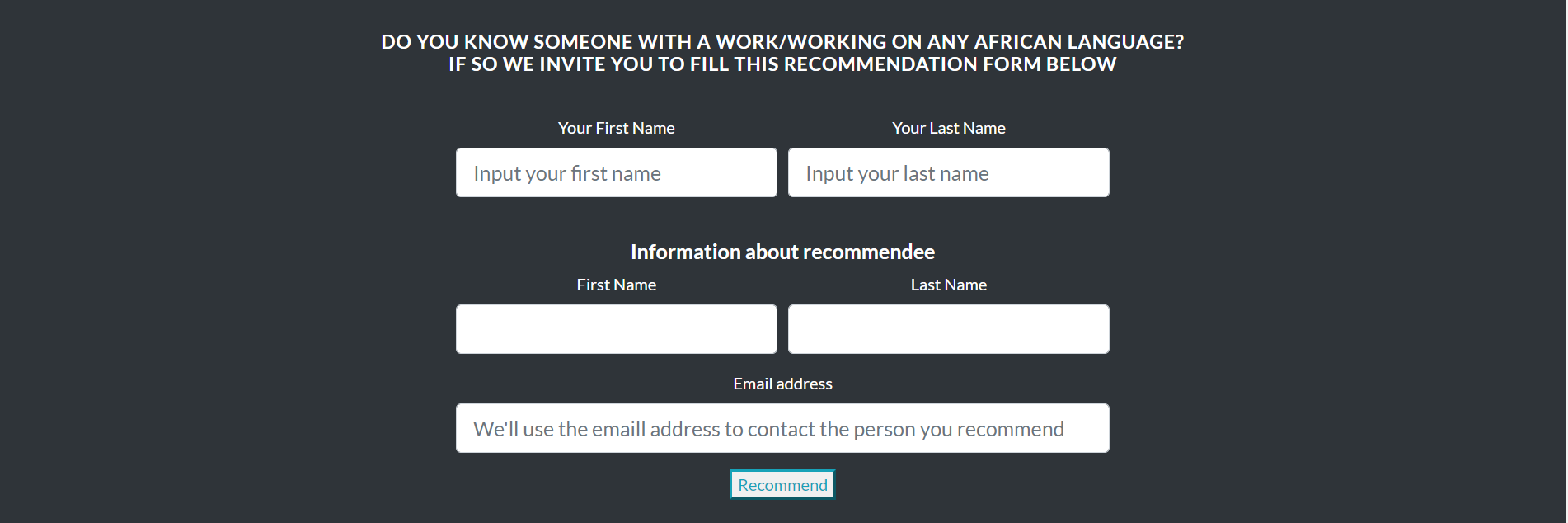}
 \caption{\small Form to recommend someone working on an African language}
\label{rec} 
\end{figure}
\item Recommend a researcher, who has an MT research paper or project on an African language, to \textit{submit a contribution} on Lanfrica as explained in step \ref{stepa}. We believe this unique feature of Lanfrica -- which connects both researchers and non-researchers in the cataloging process -- will boost networking, thereby achieving the participatory approach to documenting MT results. Now the user will be referred to a recommendation form to fill in information about the recommendee, who Lanfrica will contact, via email, to record the research on Lanfrica. See figure \ref{rec} for a screenshot of the recommendation form.
\item\label{three} View the information on the African language(s) of interest.\\
The user can, by clicking each African language, see the following tabular information on existing researches/projects (see figure \ref{resources} and \ref{data} for an example with Afrikaans):
\begin{itemize}
\item Title of research paper/project
\item Research author(s) involved: It will also be possible to contact the author(s), with the permission of the author(s), through the preferred mode of contact indicated.
\item Short description of the research paper/project.
\item Link to read/get more information about the research paper/project: to prevent copyright issues, Lanfrica will NOT host the research papers on its website, but rather refer the user (via a link) to where the paper or project can be accessed.
\item Benchmark scores (BLEU score for machine translation) for the particular African language chosen. This is an integral feature of Lanfrica: \textit{the ability to quickly get information on what benchmarks (and available test scores), if any,  exist for the African language of interest}.
\item Link for access to test data: being able to access the test/dev data will be important to compare models, thereby improving reproducibility of these researches. Therefore, if the test data used to evaluate the model is available and open-sourced, Lanfrica will put the link.
\end{itemize}
\end{enumerate}
\end{enumerate}
\section{Results}
Due to the fact that the Lanfrica web-platform is still under development, there are no results or analysis from using it yet. However, preliminary project survey, carried out to ascertain the usefulness of Lanfrica, has been summarized below:
\begin{itemize}
\item MT researchers of AfricaNLP opine that a platform like Lanfrica will ease the process of having to search around the internet just to find benchmarks on MT projects they're embarking on. They believe that Lanfrica, focusing solely on African languages, will better showcase their works to the specific community working on African languages, in contrast to platforms, like Google Scholar\footnote{\url{https://scholar.google.com/}}, which house all types of research.
\item Some organizations expressed hope for Lanfrica to partner with MT and NLP online communities in order to realize the goal of putting African NLP works on the internet. 
\end{itemize}
We also discovered that Lanfrica could serve another purpose: create data analysis on NLP researches for the different African languages, giving vital information on the level and rate of research done on each language, which could have positive unprecedented uses.

\textit{\textbf{We are currently building the web-platform for Lanfrica and plan to finish it in due time for launch during the camera-ready presentation.}}

\bibliography{example_paper}
\bibliographystyle{icml2020}

\end{document}